\documentclass[a4paper]{spie}  

\usepackage[]{graphicx}

\newcommand\memfoc{\ensuremath{f_\mathrm{m}}}

\title{Simulations of Adaptive Optics with a Laser Guide Star for SINFONI}

\author{Anthony G.A. Brown$^*$\supit{a}, Enrico Fedrigo\supit{b}, Paul van der
  Werf\supit{a}
\skiplinehalf
\supit{a}Sterrewacht Leiden, P.O.\ Box 9513, 2300 RA Leiden, The Netherlands;\\
\supit{b}European Southern Observatory, Karl-Schwarzschildstra{\ss}e 2,
  D-85748 Garching bei M\"unchen, Germany
}

\authorinfo{$^*$brown@strw.leidenuniv.nl}

\begin{document} 
\maketitle
\begin{abstract}
The SINFONI instrument for ESO's VLT combines integral field spectroscopy and
adaptive optics (AO). We discuss detailed simulations of the adaptive optics
module. These simulations are aimed at assessing the AO module performance,
specifically for operations with extended sources and a laser guide
star. Simulated point spread function (PSF) images will be used to support
scientific preparations and the development of an exposure time calculator,
while simulated wavefront sensor measurements will be used to study PSF
reconstruction methods. We explain how the adaptive optics simulations work,
focusing on the realistic modelling of the laser guide star for a curvature
wavefront sensor. The predicted performance of the AO module is discussed,
resulting in recommendations for the operation of the SINFONI AO module at the
telescope.
\end{abstract}
%
\section{Introduction}
%
The SINFONI (SINgle Far Object Near-ir Investigation) instrument is developed
jointly by the European Southern Observatory, the Max-Planck-Institut f\"ur
Extraterrestrische Physik, and the Netherlands Research School for Astronomy
(NOVA). It combines integral field spectroscopy with adaptive optics in one
instrument (see Refs.~\citenum{Bonnet2003}{ and \citenum{Eisenhauer2003}). The
instrument will be available at ESO's VLT in 2004. The AO module of the
instrument is based on a 60-element curvature wavefront sensor combined with a
bimorph deformable mirror. The wavefront sensor employs 60 avalanche
photo-diodes (APDs) as detectors. The AO system will operate in natural (NGS)
and laser guide star (LGS) mode and will feed the corrected PSF to the
near-infrared integral field spectrograph SPIFFI\cite{Eisenhauer2003}.

As part of the NOVA contribution to this project detailed simulations of an AO
system with a curvature sensor, tailored to the details of the SINFONI AO
module, are being carried out. The aim is to develop a realistic simulation of
the operation of the AO module with a laser guide star generated in the
mesospheric sodium layer. The output consists of detailed statistics of the AO
system performance as well as PSF images (at $J$, $H$, and $K$-band) and
wavefront sensor data. The results will be used to support the development of
algorithms for PSF reconstruction from wavefront sensor data (see
Ref.~\citenum{Rigal2004} in this volume) and the development of an exposure
time calculator. Scientific preparations for the use of the SINFONI instrument
will also be supported with these simulations.
\begin{figure}
\begin{center}
\includegraphics[height=0.9\textheight]{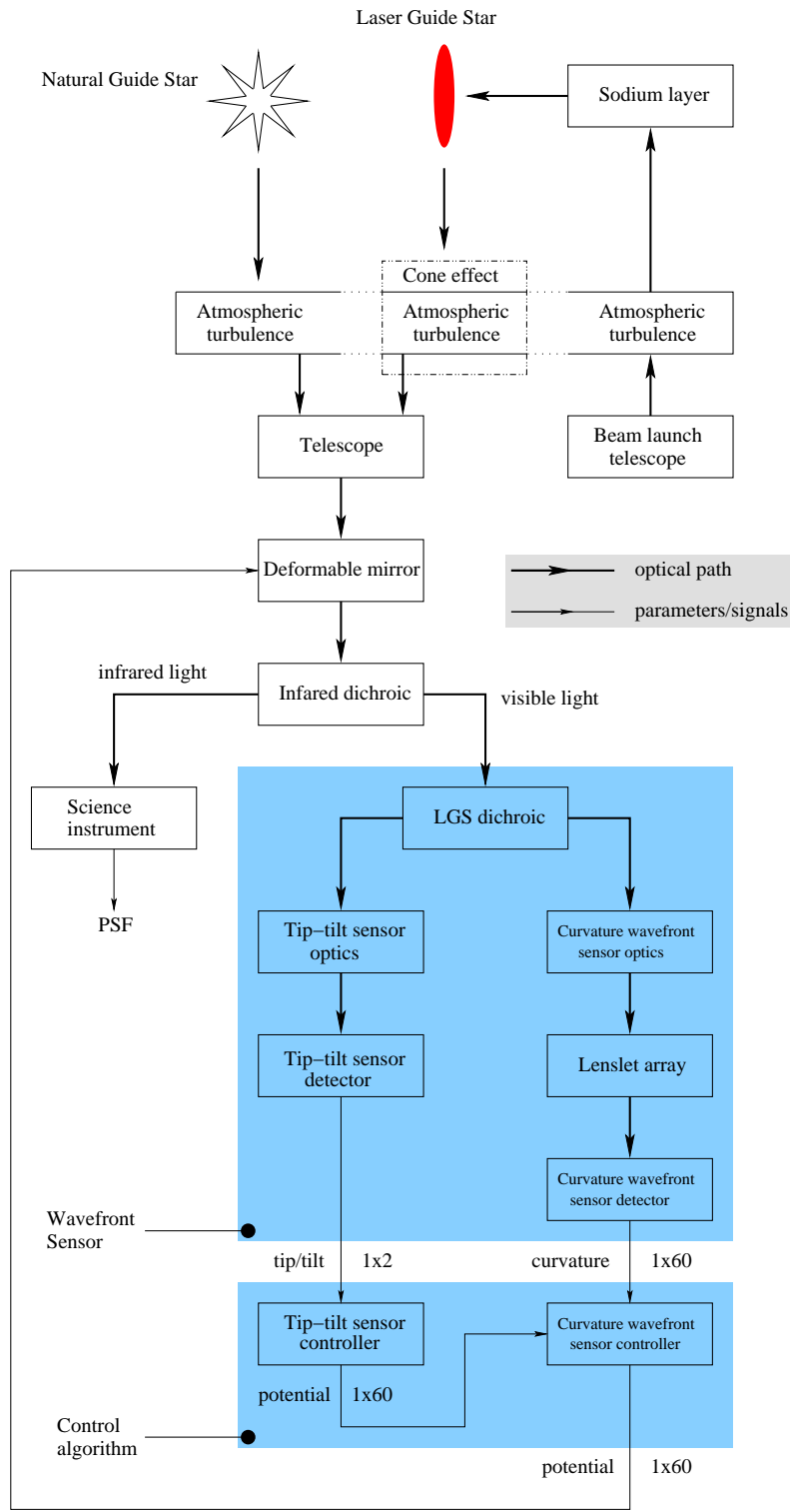}
\end{center}
\caption{\label{fig:ao-system} Structure of the AO simulation package. The
figure is after Ref.~\protect\citenum{Craven2000} (Fig.\ 4.1) with the
addition of the laser guide star parts. This includes the cone effect for the
LGS and the presence of a natural guide star for tip-tilt sensing.}
\end{figure}
%
\section{Overview of the basic SINFONI AO simulation tool}
%
The simulation code is based on a Matlab adaptive optics plus curvature
wavefront sensor simulation package developed at ESO\cite{Craven2000}. The
basic structure of the simulated AO system as well as the light and signal
paths are shown in Fig.~\ref{fig:ao-system}. The atmospheric turbulence is
simulated with three phase screens to which a height, weight, and wind speed
can be assigned. The phase structure function simulated by the phase screens
follows a Kolmogorov spectrum and the screens are normalised such that unit
phase variance (1 radian$^2$) is attained over a circular region of 128
samples in diameter. The screens can thus be normalised for any value of the
Fried parameter $r_0$ by scaling with $\sqrt{1.03(D/r_0)^{5/3}}$, where $D$ is
the diameter of the telescope primary mirror, which is sampled by 128
samples. The simulations discussed below use a value of $r_0=16$~cm, which
corresponds to a seeing of $0.65$ arcsec at an observing wavelength of 550 nm.

The AO corrections are done with a deformable mirror (DM) which is simulated
through a detailed finite-element model of the bimorph mirror that will be
used in the real system. The curvature wavefront sensor optics module
implements the membrane mirror as a time-varying focus term that is added to
the phase of the optical field. The lenslet array module divides the light
among the 60 elements of this array and the light is then detected with
simulated APDs by the detector module. The simulations include the variation
of the field of view of the lenslet array with the membrane mirror focal
length. This is implemented as a vignetting of the optical field after
reflection off the membrane mirror. The curvature signal is constructed as the
difference of the accumulated intra- and extra-focal intensity distributions
and sent to the controller module which steers the DM. For LGS operation an
additional tip-tilt (TT) sensor is needed which is modelled as a simple
quad-cell.

For propagating the light through the system the Fraunhofer approximation is
used everywhere, which implies a Fourier transform at each propagation from a
pupil to an image plane or vice versa.

At the start of the simulation the interaction matrix between the curvature
sensor and the DM is determined by applying voltages to each actuator of the
DM in turn and then measuring the resulting curvature signal. The control
matrix is the pseudo inverse of the interaction matrix, which is obtained
after filtering out the smallest eigenvectors from the latter.

The basic AO simulation package developed at ESO only handled the case of a
natural guide star (i.e.\ a point source outside the atmosphere) and was
extended to include a laser guide star. The beacon produced by the laser guide
star facility in the sodium layer will be a three-dimensional region from
which photons are returned to the telescope. The AO simulation thus has to
model how the beacon is generated and how the curvature sensor `sees' a 3D
light distribution, extended along the line of sight. In the case of the VLT
the laser light will be launched from behind the secondary mirror which means
that the laser beacon is always on axis. This allows the modelling of the
beacon as a stack of extended (2D) sources (see Sect.~\ref{sec:lgs}). The
first extension of the AO simulation tool was thus to include extended
(natural) guide sources which is described in the following section. This has
an interest of its own as in reality there are many occasions where one may
want to guide the AO system on an extended source. In Sect.~\ref{sec:lgs} a
detailed description is given of the subsequent implementation of the laser
guide star.
\begin{table}[t]
\caption{Parameters of the AO simulations with a natural guide source which
  can be a point source or extended. The column on the left lists the
  parameters that can be varied during the simulations and the column on the
  right shows the default value, or range of values, used in the simulations.}
\label{tab:egsparams}
\begin{center}
\begin{tabular}{l|l}
\multicolumn{1}{c}{Model parameter} & \multicolumn{1}{c}{Value} \\
\hline
\multicolumn{2}{c}{Overall} \\
\hline
Seeing & 0.65 arcsec at 500 nm\\
NGS/EGS wavelength & 700 nm \\
NGS/EGS magnitude & 9--16 \\
NGS/EGS size & 0--2 arcsec FWHM (Gaussian light distribution)\\
Sky background magnitude & 19 \\
Science wavelength & $J$, $H$, $K$ \\
Integration time & 1 second \\
Position of NGS/EGS and science target & Both on-axis and at zenith\\
Telescope focal length & 374.4~m\\
\hline
\multicolumn{2}{c}{Wavefront sensor} \\
\hline
Membrane mirror focal length & 10--100~cm\\
Membrane mirror frequency & 2100~Hz\\
Time step & 29.8 micro-seconds (16 samples per membrane period)\\
Atmosphere update interval & 238 micro-seconds (8 time steps)\\
Curvature sensor loop gain & 0.1--0.9\\
Curvature sensor integration periods & 6\\
APD read-out delay & 0 seconds\\
APD read-out noise & 0 e$^{-}$\\
APD curvature sensor quantum eff. & 0.70\\
APD curvature sensor dark current & 250 e$^{-}$/sec\\
APD tip-tilt sensor quantum eff. & 0.70\\
APD tip-tilt sensor dark current & 250 e$^{-}$/sec\\
\end{tabular}
\end{center}
\end{table}
%
\section{Implementing an extended natural guide source}
\label{sec:egs}
%
The extended guide source (EGS) was implemented in the simulations by
considering it to be a collection of independent point sources at different
positions on the sky that all contribute to the extended source signal. It is
assumed that the light from each point source travels along the same path
through the atmosphere (i.e., anisoplanatism effects over the extent of the
source are neglected). The light of these point sources will travel from the
VLT primary mirror to the curvature wavefront sensor and there reflect off the
membrane mirror which is located in the focal plane of the telescope. After
reflection the light goes on to the lenslet array where for a flat membrane a
replica of the VLT entrance pupil (the pupil image) is formed.

For a flat membrane mirror the pupil image at the lenslet array will be a
uniformly illuminated disk (with a hole in the middle caused by the VLT
secondary). When the membrane mirror is curved a defocus term is added to the
optical field reflecting off the mirror which results in a pupil image at the
lenslet array that contains information on the phase structure of the optical
field at the VLT primary (this is the basis for the functioning of a curvature
sensor). In the presence of a pure tilt in the phase of the optical field at
the VLT pupil, the defocus term leads to a shift of the pupil image on the
lenslet array.

This means that for a collection of point sources on the sky the resulting
image will be the convolution of the pupil image for an on-axis point source
with the distribution of point sources (i.e.\ the extended source intensity
distribution) on the sky. In order to implement this correctly in the
simulations it is necessary to know how the angular scale on the sky
translates to the linear scale at the lenslet array. This can be derived from
a Fourier analysis of the path that the light takes from the VLT pupil to the
lenslet array. It can be shown that in the presence of a pure tilt in the
wavefront the pupil image only undergoes a shift $\Delta s$ (in mm) which is
given by:
\begin{equation}
\Delta s=\tau\frac{(f-\memfoc)}{\memfoc}f_\mathrm{L}\approx
\tau\frac{ff_\mathrm{L}}{\memfoc}=\frac{\zeta}{\memfoc}\tau\,,
\end{equation}
where $\tau$ is the angular distance from the optical axis (in arcseconds on
the sky), $f$ is the focal length of the beam coming from the primary mirror,
{\memfoc} is the membrane mirror focal length\footnote{The focal length of the
membrane mirror will be used throughout this paper instead of the more common
radius of curvature (which is twice the focal length).}, and $f_\mathrm{L}$ is
the focal length of the beam travelling from the membrane mirror to the
lenslet array. The approximation on the right is valid to within 2 per cent
for the range of values of {\memfoc} occurring in practice. In the simulations
we use $\zeta=1220$~mm$^2$/arcsec ({\memfoc} in mm and $\tau$ in arcsec).
\begin{figure}
\begin{center}
\includegraphics[height=0.6\textwidth]{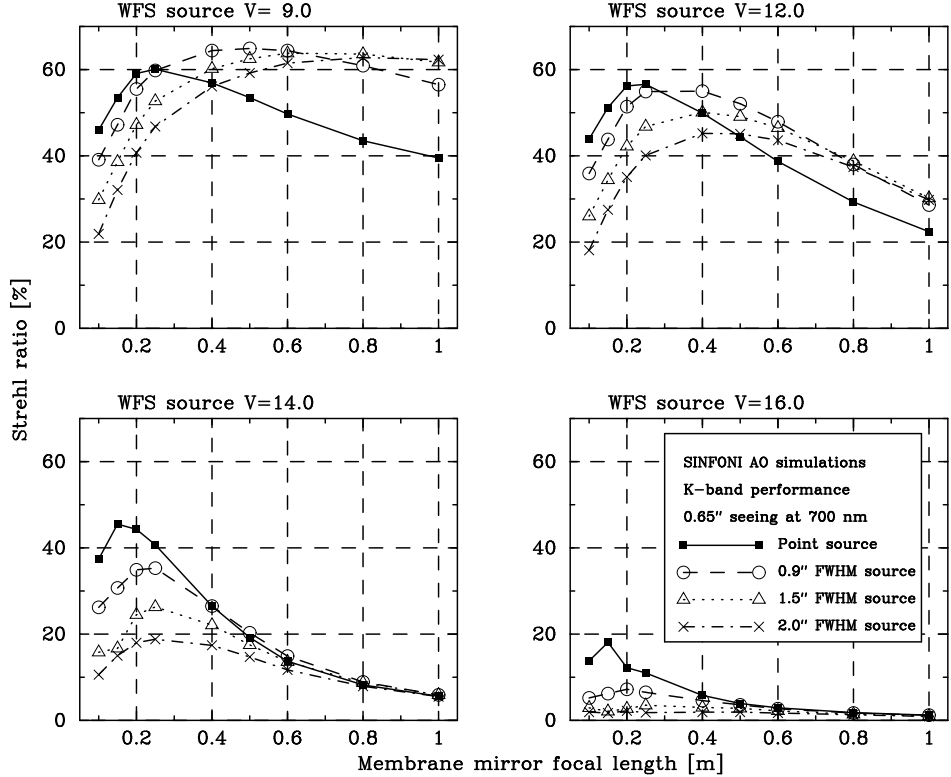}
\end{center}
\caption{\label{fig:egs-strehls-k} The four diagrams show the performance of
  the SINFONI AO module for natural guide sources at $K$-band. The light
  distribution of the extended sources is Gaussian. The Strehl ratio of the
  PSF is measured after 1 second of integration time. The four panels are for
  different magnitudes of the guide source ($V=9\,, 12\,, 14\,, 16$) and the
  Strehl ratios are shown for different source sizes (0, 0.9, 1.5, and 2.0
  arcsec FWHM; as measured on the sky outside the atmosphere) as a function of
  the focal length of the membrane mirror of the curvature sensor.}
\end{figure}
\begin{figure}
\begin{center}
\includegraphics[height=0.6\textwidth]{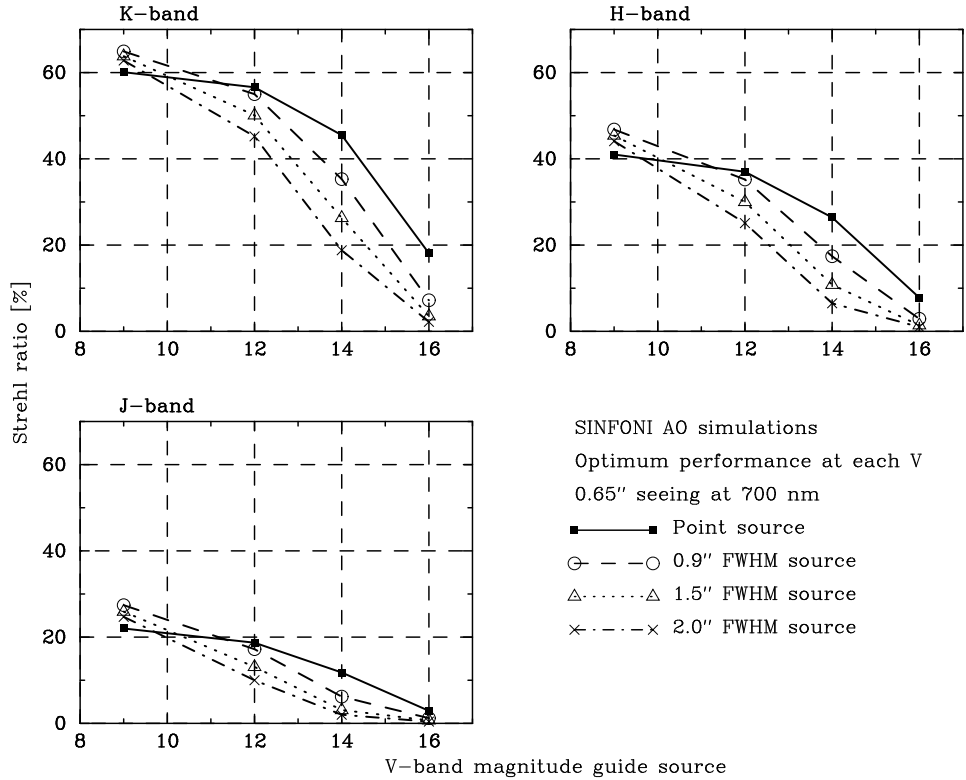}
\end{center}
\caption{\label{fig:egs-optimum-strehls} The performance of the AO system for
  the $J$, $H$, and $K$-bands as measured by the Strehl ratio. The performance
  at each magnitude was optimised by varying the membrane mirror focal
  length (see Fig.~\ref{fig:egs-strehls-k}).}
\end{figure}
\begin{figure}
\begin{center}
\includegraphics[height=0.6\textwidth]{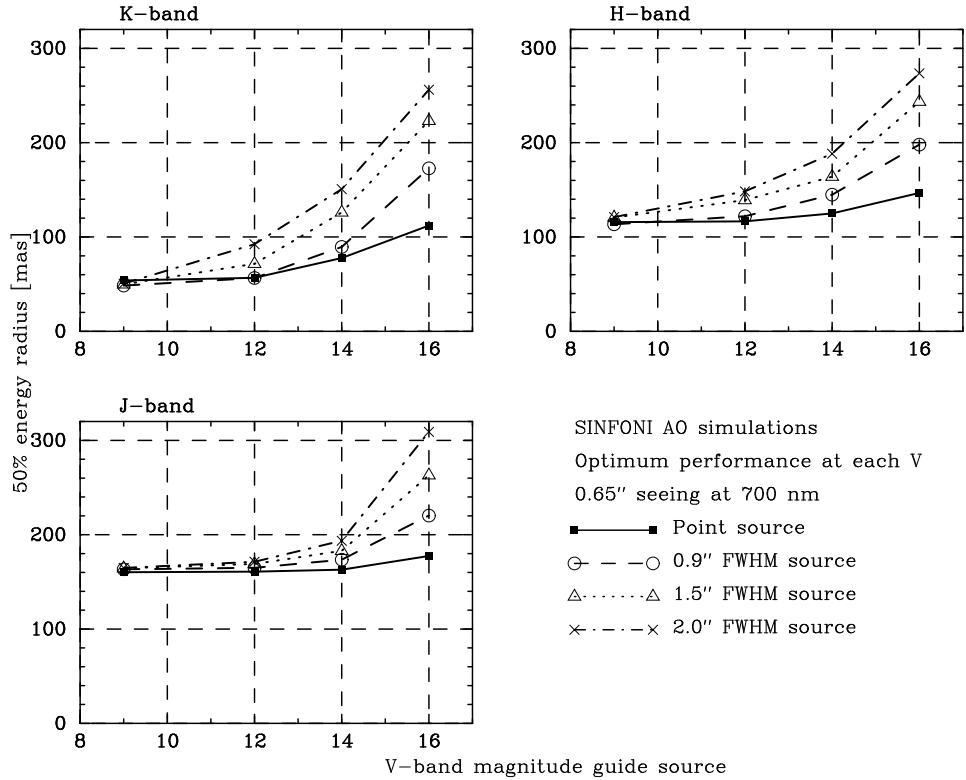}
\end{center}
\caption{\label{fig:egs-optimum-en50} Optimum performance of the AO system at
  $J$, $H$, and $K$ as measured by the 50\% encircled energy radius (in units
  of milli-arcseconds).}
\end{figure}
%
\subsection{Predicted performance for natural guide sources}
\label{sec:egsperf}
%
Having implemented an extended natural guide source in the simulations
we can now use them to predict the performance on such sources and compare
that to the performance for point sources. The extended sources can in reality
have arbitrary intensity distributions on the sky but we only investigated
Gaussian light distributions for a range of FWHM values. The parameters of the
simulation that can be tuned are listed in Table~\ref{tab:egsparams}.

The simulations were used to investigate how the performance of the AO system,
as measured by the Strehl ratio of the PSF, varies as a function of the
brightness and size of the guide source, the membrane mirror focal length and
the curvature sensor control-loop gain. The results can be used during actual
operations of the AO system to guide the optimisation of the settings.

The resulting performance figures for a control loop gain of $0.6$ (which is
close to optimal in most cases) are shown for the $J$, $H$, and $K$ bands in
Figs.~\ref{fig:egs-strehls-k}--\ref{fig:egs-optimum-en50}. Figure
\ref{fig:egs-strehls-k} shows how the performance varies as a function of the
membrane mirror focal length for various source sizes and magnitudes. Figures
\ref{fig:egs-optimum-strehls} and \ref{fig:egs-optimum-en50} show the Strehl
ratio and 50\% encircled energy radius at each magnitude after the membrane
mirror focal length has been adjusted to its optimal value.

From considerations of the characteristics of curvature based systems (see
Ref.\ \citenum{Rousset1999}, Sect.\ 5.3.3) it is expected that for extended
guide sources one should increase the extra-focal distance realised with the
membrane mirror (i.e., increase the mirror focal length or, equivalently,
reduce the stroke of the membrane) in order to optimise the
performance. However, the sensitivity of the system decreases with larger
extra-focal distances so some comprise between membrane stroke and source
brightness is expected to optimise the performance for the EGS case.

This is confirmed by the results obtained from the simulation of the SINFONI
AO module. Inspection of Fig.~\ref{fig:egs-strehls-k} shows a general trend
that for a given source magnitude (for the brighter sources) the performance
reaches its optimum for longer extra-focal distances as the source size
increases. On the other hand, for a given guide source size the optimum
performance shifts toward smaller extra-focal distances as the source
brightness decreases (the same behaviour is seen for $J$ and $H$).

However for the brightest guide sources the optimum performance for the
extended sources is always better than the optimum performance for the point
source (as judged by the Strehl ratio and the 50\% encircled energy
radius). Intuitively one expects that no matter how well optimised the EGS
case is, information is always lost due to the smearing out of the
signal. However this smoothing of the signal also leads to the filtering out
of the high frequency spatial aliasing that occurs in curvature wavefront
sensors. This improves the performance at high signal-to-noise ratios
(F.~Rigaut, priv.\ communication).

For operating the SINFONI AO module on natural guide sources the following
recommendations can be made based on the simulation results: For point sources
brighter than $\textit{V}\sim\textrm{14}$ the optimum membrane mirror focal
length is $\sim0.25$~m. This value should be lowered to $0.1$--$0.2$~m for
fainter sources. For extended sources brighter than
$\textit{V}\sim\textrm{14}$ the membrane mirror focal length should be
adjusted during the operation of the AO system in order to improve
performance. For extended guide sources fainter than
$\textit{V}\sim\textrm{14}$ the membrane mirror focal length should be kept at
about 0.25~m. Experiments with the simulations show that a re-calibration of
the interaction matrix for extended sources (by calibrating on an artificial
extended source) does not improve the performance and thus is not
necessary. The calibration can always be done on a point source.
%
\section{Implementing the laser guide star}
\label{sec:lgs}
%
The laser guide star facility that will be installed at ESO's VLT will make
use of the mesospheric sodium layer to produce the laser beacon. The telescope
that will launch the laser light will be located behind the secondary mirror
of the VLT, resulting in a monostatic projection (i.e.\ on-axis with respect
to the primary mirror)\cite{Bonaccini2003}. Figure~\ref{fig:lgs-simulation}
shows a schematic diagram of this situation. The laser beam is launched by the
beam launch telescope (BLT) and travels upwards to the sodium layer, spreads
out along the way due to diffraction, and suffers the effects of
turbulence. In the sodium layer the laser light is resonantly back-scattered
and this results in a laser beacon which because of the finite width of the
layer is extended in 3 dimensions\footnote{Note: the 3D distribution of
back-scattered laser light in the sodium layer will be referred to as the
laser `beacon', whereas the image of this beacon as seen from the ground will
be referred to as the laser `spot'.}. Subsequently the laser light travels
back down to the telescope, again going through turbulence, whereafter it will
be analysed by the AO system. Turbulent layers at different heights are
present and the sodium layer is located a height $z_0$ and has a thickness
$L$.

Figure~\ref{fig:lgs-simulation} also shows how the laser beacon is actually
modelled in the simulation. It is represented as a stack of 2D extended
sources. For each of these one needs to know the height, the number of photons
returned, and the distribution of photons in the 2D spot before downward
propagation of the laser light. The laser beam optical field at launch is
assumed to have a constant phase which upon propagation to the sodium layer
will be affected by turbulence. The resulting intensity distribution in the
sodium layer corresponds to the image of the 2D sources unaffected by
turbulence from the downward path. Finally, the light from each 2D source will
propagate back to the telescope and pass through turbulence layers and now the
cone effect and loss of tip-tilt information in the laser signal have to be
taken into account.
\begin{figure}[t]
\begin{center}
\includegraphics[height=0.55\textheight]{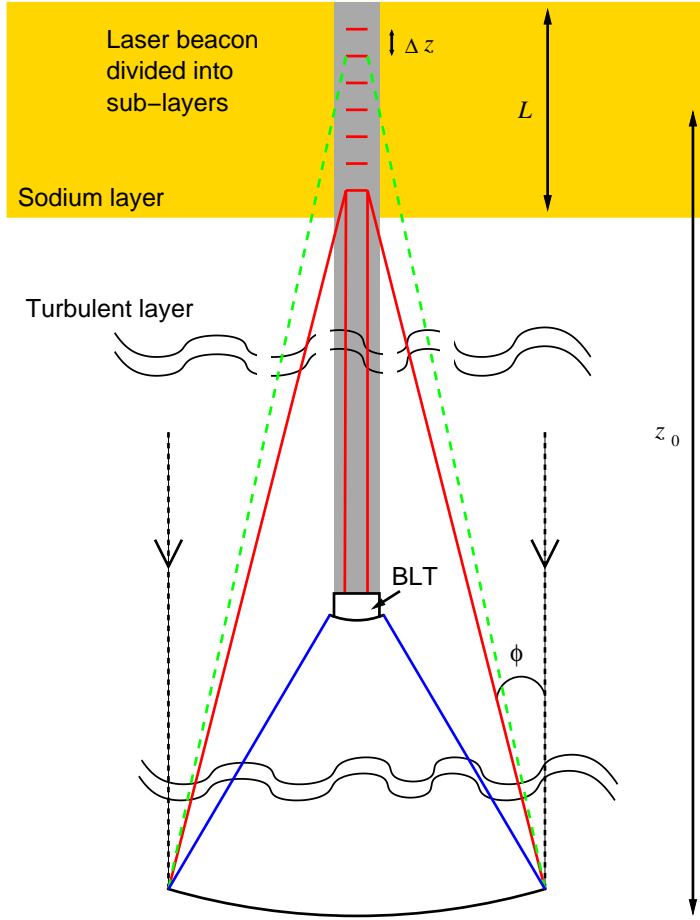}
\end{center}
\caption{\label{fig:lgs-simulation} Schematic diagram of the simulation of a
laser guide star for SINFONI. The sodium layer is divided into sub-layers that
are considered to be so thin that the resulting back-scattered photons come
from a 2D region. Thus the LGS is simulated as a stack of extended 2D guide
sources. For each EGS one has to calculate its light distribution and the
location where it will be focused. The response of the curvature sensor to the
separate EGS's can then be summed to obtain the LGS response. The differential
cone effect, illustrated here by the two cones coming down from different
layers in the laser beacon, can be ignored in the SINFONI simulations.}
\end{figure}
%
\subsection{Laser beam launch and upward propagation}
\label{sec:beamlaunch}
%
The beam launch telescope will have a primary mirror of 50 cm (clear aperture
outer diameter) and a central obscuration of $4.2$~cm. The profile of the
laser beam can be taken to be a Gaussian (i.e., the optical field has an
amplitude which has a Gaussian shape perpendicular to the beam direction). The
launched $1/e^2$ diameter of the laser beam will be 35
cm\cite{Bonaccini2003}. The BLT is assumed to be focused at the altitude of
the centroid of the sodium layer and then the irradiance distribution at that
point due to the laser light can be obtained by a Fourier transform of the
laser optical field. The FWHM size of this spot in the absence of turbulence
will be about $0.29$ arcseconds for the VLT laser guide star facility. The
presence of atmospheric turbulence on the way up to the sodium layer is taken
into account by adding the phase from the atmospheric phase screens to the
optical field of the laser beam. The atmospheric phase screens are represented
by a $128\times128$ matrix over the size of the VLT pupil (8~m) and the inner
$8\times8$ samples (the $50\times50$~cm$^2$ region seen by the BLT) of the
phase screens are used to simulate the effect of turbulence on the laser
beacon. The cone effect on the upward path is ignored in the simulations.

After Fourier transforming the laser optical field including the turbulence
one obtains the intensity distribution of the 2D sources that represent the
laser spot. Just as for the extended natural guide sources
(Sect.~\ref{sec:egs}) this intensity distribution is convolved with the point
source pupil image at the lenslet array of the curvature sensor. In this way
the 2D extent of the laser guide star across the line of sight is taken into
account.

The diameter of the BLT spans only a few turbulence `cells' of size $r_0$
(i.e., $D/r_0\sim \textrm{few}$). This implies a high Strehl ratio for short
exposure images of the laser spot, making the spot size not much larger than
the $0.29$ arcseconds quoted above. Note that this refers to the spot size
\textit{before} downward propagation.
%
\subsection{Laser beacon generation}
%
This part of the simulation code takes care of generating a realistic
distribution of the returned photons as a function of the location within the
3D laser beacon. The theory of photon return from sodium beacons is
extensively discussed in two papers by Milonni and co-workers
\cite{Milonni1998,Milonni1999} and their results are used in the calculations
of the photon return rate in the simulation. For a continuous-wave laser as
used for the VLT the photon return per unit area at the receiver can be
written as:
\begin{equation}
\label{eq:photreturn}
R_\mathrm{cw}\propto\frac{T_0C_\mathrm{S}}{z_0^2}\,,
\end{equation}
where the constants are the atmospheric transmission $T_0$, the sodium atom
column density $C_\mathrm{S}$, and the height of the sodium layer $z_0$. Note
that the atmospheric transmission only contains the downward component. In
reality one deals with an overall transmission $\eta T_0^2$, where $T_0$
occurs twice to account for the up and down propagation of the laser beam and
the factor $\eta$ includes the transmission coefficients of the beam
transportation system, the beam launch telescope, and the astronomical
(VLT-UT) telescope.

When dividing the laser beacon into discrete 2D sources
Eq.(\ref{eq:photreturn}) can be used to calculate for each 2D source the
relative number of photons (with respect to the sodium centroid altitude)
returned to the curvature sensor. Dividing the sodium layer into $n$ discrete
layers of thickness $\Delta z$, the column density can be obtained by
integrating of the vertical sodium density distribution $\rho(z)$ and the
atmospheric transmission can be calculated relative to the transmission
corresponding to the sodium centroid altitude. The number of photons
$N_{\mathrm{ph},i}$ returned from each sub-layer $i$ is given by:
\begin{equation}
\label{eq:naweights}
N_{\mathrm{ph},i}=N_\mathrm{ph}\frac{w_i}{\sum_i w_i} \quad
\textrm{and}\quad
w_i=\left(\frac{z_0}{z_i}\right)^2%
e^{\frac{2\ln T_0}{z_0}(z_i-z_0)}%
\int_{z_i-\frac{\Delta z}{2}}^{z_i+\frac{\Delta z}{2}}\rho(z)\mathrm{d}z\,,
\end{equation}
where $N_\mathrm{ph}$ is the total number of photons (summed over the
sub-layers), $z_i$ is the centroid altitude of sub-layer $i$, which has a
thickness $\Delta z$. The first factor in the calculation of $w_i$ takes the
geometrical dilution of the photon return into account, the second factor
describes the atmospheric transmission, and the third factor is the column
density for sub-layer $i$.  The relative transmission is calculated assuming
that the optical depth is proportional to $z$. The factor 2 in the exponential
takes into account the upward and downward path of the laser light through the
atmosphere (the extra transmission factor $\eta$ mentioned above is the same
for all sub-layers and need not be explicitly known for this calculation). The
density $\rho(z)$ is assumed to be a Gaussian with a full width half maximum
equal to the thickness $L$ of the sodium layer. Note that the geometrical
dilution and transmission factors should be included in the integral but have
been approximated here by their values at the distance $z_i$ of each sub-layer
$i$.
\begin{table}[t]
\caption{Parameters of the AO simulations with a laser guide star. These
  parameters are used in addition to those listed in Table
  \protect\ref{tab:egsparams}. The column on the left lists the parameters
  that can be varied during the simulations and the column on the right shows
  the default value, or range of values, used in the simulations. (SCA stands
  for sodium centroid altitude.)}
\label{tab:lgsparams}
\begin{center}
\begin{tabular}{l|l}
\multicolumn{1}{c}{Model parameter} & \multicolumn{1}{c}{Value} \\
\hline
\multicolumn{2}{c}{Laser beacon and tip-tilt star}\\
\hline
LGS wavelength & 589 nm \\
NGS (tip-tilt star) wavelength & 700 nm \\
Position of LGS beacon and science target & Both on-axis\\
Return flux from LGS & $10^6$ photons/sec/m$^2$ \\
Atmospheric transmission up to SCA & 0.7\\
Laser beam launch & On axis\\
Beam launch telescope diameter & 50 cm; focused to SCA \\
Laser beam & Gaussian; $1/e^2$-diameter 35~cm \\
Sodium layer & Height 90~km asl, thickness 3--10~km, Gaussian density distrib. \\
No.\ of layers in sodium beacon model & 3\\
Telescope altitude & 2635~m asl \\
SCA measurement error & 0--few hundred m\\
LGS zenith angle & 0--60~degrees\\
TT (tip-tilt) star magnitude & 12--18\\
TT star offset from LGS & 0--60~arcsec\\
TT star position angle wrt LGS & 0--360~degrees\\
\hline
\multicolumn{2}{c}{Control loops} \\
\hline
Tip-tilt sensor integration periods & 6\\
TT sensor loop gain & 9 \\
BLT steering mirror loop gain & 2.2 \\
BLT steering mirror time delay & 0--$N$ time-steps \\
\hline
\multicolumn{2}{c}{Atmosphere} \\
\hline
Layer heights & 0, 7000, 13000 meter\\
Layer weights & 0.6, 0.2, 0.2\\
Layer speeds & 5.7, 5.7, 33.0 meter/sec\\
\end{tabular}
\end{center}
\end{table}
\begin{figure}
\begin{center}
\includegraphics[width=0.8\textwidth]{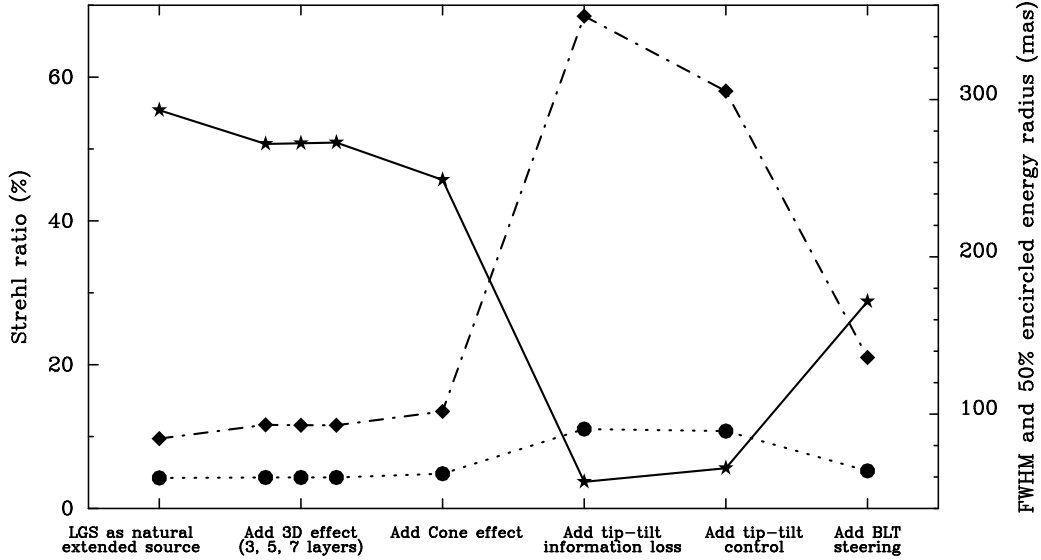}
\end{center}
\caption{\label{fig:addrealism} The effects of increasing the realism of the
laser guide star simulations. The plot shows the Strehl ratio (starred
symbols, left vertical scale), the full-width half-maximum and the 50\%
encircled energy radius (in milli-arcseconds, filled circles and diamonds
respectively, right vertical scale) of the PSF at K-band after 1 second of
integration time. From left to right the realism of the simulations is
increased by adding the effects listed on the horizontal axis.}
\end{figure}
%
\subsection{Downward propagation and the cone effect}
%
On downward propagation of the laser light the atmosphere will not be fully
sampled. The atmosphere above the LGS is not seen at all and below the LGS
only the atmosphere within a cone of opening angle $2\phi$ will be seen (see
Fig.~\ref{fig:lgs-simulation}). For an 8~m telescope and a height of the LGS
of 90~km, $2\phi\sim9\times10^{-5}$ rad. This means for example that of a
turbulent layer at a height of 10~km only a circle of 7.1~m diameter will be
seen. There is also a differential cone effect due to the extent of the laser
beacon along the line of sight (see Fig.~\ref{fig:lgs-simulation}). This means
that light coming from the bottom and top of the laser beacon arriving at the
same position on the telescope primary mirror will have travelled at slightly
different angles through the turbulence, thus causing a differential
anisoplanatism error. For the SINFONI case this can be ignored as the
additional information smearing occurs on scales of a few tens of centimetres
which is much smaller than the resolution elements of the curvature sensor
across the primary mirror (which are of order 1 meter).

For downward propagation the cone effect is taken into account by calculating
for each turbulent atmosphere layer what fraction of the $8\times8$~m phase
screen ($128\times128$ samples across the VLT pupil) is seen from the sodium
layer. The resulting $n\times n$ ($n<128$) sample phase screen is scaled to
$128\times128$ samples. All the resulting phase screens from each turbulent
layer are then added to obtain the overall effect on the laser beacon. The
response at the curvature sensor is calculated for each sub-layer making use
of the 2D intensity distribution which is obtained as explained in
Sect.~\ref{sec:beamlaunch}.
%
\subsection{3D effect at the curvature sensor}
%
The different 2D sources in the stack that simulates the laser beacon are not
all located at the sodium centroid altitude to which the AO system is
focused. This means that the layers away from the sodium centroid altitude
will be seen out of focus at the curvature sensor. Equivalently one can think
in terms of the outer parts of the curvature sensor lenslet array seeing an
elongated laser spot. The effect of the 3D extent of the laser beacon is a
radial smearing of the information at the lenslet array. This effect is
implemented as an additional focus term that is added to the optical phase for
each sub-layer $i$, which is proportional to
$(1/z_i-1/z_0)\times(\xi^2+\eta^2)$, where $(\xi,\eta)$ are the coordinates
across the VLT pupil. After the response of the curvature sensor to each
individual sub-layer $i$ is calculated all the signals are added to obtain the
overall response.
%
\subsection{Tip-tilt management}
%
One of the most important issues when using laser guide stars is that of the
tip-tilt corrections to be applied to the wavefront corresponding to the
scientific target. The laser light launched from the BLT will suffer the
effects of turbulence on both the upward and downward paths leading to a
summed tip-tilt component in the laser light phase that is different from and
uncorrelated with the tip-tilt component in the phase of the light from the
science target. This necessitates the addition of a natural guide star for
tip-tilt sensing (the `tip-tilt star') in order to correct the tip-tilt of the
science target wavefront. The tip-tilt star will in general be located away
from the science target which leads to tilt-anisoplanatism errors in the
corrected wavefront. We now describe how these issues are accounted for in the
simulations.

In the simulations the tip-tilt components of the atmospheric phase screen
seen by the BLT are saved and then subtracted from the phase screen to be used
for downward propagation (after multiplication of the tip-tilt coefficients by
the ratio of the VLT to BLT pupil). This correctly simulates how the tip-tilt
motion of the laser beacon is affected by up- and down propagation of the
laser light.

A separate tip-tilt sensor is implemented in the simulations as a simple
quad-cell detector which senses the wavefront from a natural guide star. In
the SINFONI AO module the tip-tilt compensation of the science wavefront will
be done with the deformable mirror which is inserted in a tip-tilt mount. The
deformable mirror will thus be steered by both the curvature and the tip-tilt
sensor so one has to take care in not having the tip-tilt signal contained in
the laser light interfere with the tip-tilt steering. This is implemented in
the simulation by removing from the curvature controller potential vector the
part corresponding to a flat mirror surface and replacing that by the
potential vector from the tip-tilt controller.

The tip-tilt signal at the curvature sensor due to the laser guide star will
be large if the position of the LGS on the sky is not stabilised. This will
then drive the curvature signal into the non-linear regime and render the
higher order corrections invalid. Thus the LGS itself also has to be
stabilised which will be done by a steering mirror inserted into the light
path at beam launch. The required tip-tilt compensation is contained in the
tip-tilt signal measured in the laser light by the curvature sensor. The
latter signal is used in the simulations to control the steering mirror of the
BLT.
%
\subsection{Additional simulation elements}
%
A number of options have been added to the simulation package to facilitate
the study of the AO system performance under varying observing conditions. The
height and thickness of the sodium layer can be varied and a constant focus
error can be added (which simulates a telescope focused at the wrong
height). In addition observations can be simulated away from zenith and the
natural guide star for tip-tilt sensing can be placed off-axis to simulate
tilt-anisoplanatism effects. All parameters of the LGS components of the
simulation are listed in Table~\ref{tab:lgsparams}.
%
\section{LGS simulation results}
%
After putting together all the LGS simulation elements described in the
previous section one can investigate the performance predicted for the SINFONI
AO system used with a laser guide star. Figure~\ref{fig:addrealism} shows the
performance of the simulated SINFONI AO system as the realism of the
simulations is increased. The membrane mirror focal length was 25 cm in these
simulations. The three lines show for 1 second of integration time; the Strehl
ratio (line with starred symbols, scale on the left), the FWHM (line with
dots, scale on the right) and the 50\% encircled energy radius (line with
diamonds, scale on the right). The leftmost points show the performance if one
would take the 2D laser spot corresponding to an infinitesimally thin sodium
layer and put it outside the atmosphere. Thus the performance with respect to
a point source is only degraded by the source extent on the sky. From left to
right the realism of the simulations is increased by adding first the
3D-effect (using 3, 5, and 7 layers for the laser beacon), then the
cone-effect, and finally the effects of the loss of tip-tilt information in
the laser light.

These results show that the 3D light distribution of the LGS beacon is already
well modelled by three sub-layers. The combination of the 3D and cone effects
lead to a $\sim10$\% loss in Strehl ratio (starting from 55\%). Note the
dramatic performance loss once the tip-tilt complications are included. The
addition of a tip-tilt star does not improve the performance unless at the
same time the launched laser beam is stabilised on the sky. The performance
for the latter case corresponds to a Strehl ratio of 29\%. This can be further
optimised by changing the membrane mirror focal length to 40 cm and tuning the
gains of the three control loops. The optimum performance achieved is then
36\% at K-band.
%
\section{Conclusions and future work}
%
We have described detailed simulations of the SINFONI AO module and used these
to make performance predictions for operation with natural (extended) guide
sources and a laser guide star. The results can be used to guide the
optimisation of the AO module settings during actual observations (see
Sect.~\ref{sec:egsperf}) and to support other activities such as the study of
the reconstruction of the PSF from wavefront sensor data and the scientific
preparation of observations with SINFONI.

Concerning the laser guide star mode a number of tasks and open issues
remain. The predicted performance of the laser guide star mode of the SINFONI
AO module will be mapped as a function of various parameters, such as the
tip-tilt star brightness and distance from the science target, the zenith
angle, the characteristics of the sodium layer, and the natural seeing. The
steering of the BLT mirror is currently done in open loop in the simulations
and is quite sensitive to the exact gain settings for the three control loops
present in the AO module. The LGS performance can possibly be improved further
by switching to closed loop control of the BLT steering mirror and further
investigating how the different control loops interact and how their gains
should be optimised.
%
\acknowledgements
%
We thank Rudolf Le Poole, Miska Le Louarn, Enrico Marchetti and Markus Kasper
for fruitful discussions that greatly helped along the development of the
simulations at various stages.
%
\bibliography{paper}   
\bibliographystyle{spiebib}   
\end{document}